# Leaky-wave Coil Element with Improved Tx-efficiency for 7T MRI Using a Non-Uniform Current Design

K. Popova, R. Balafendiev, J. T. Svejda, *Member, IEEE*, A. Rennings, *Member, IEEE*, A. J. Raaijmakers, C. M. Collins, R. Lattanzi, D. Erni, *Member, IEEE*, G. Solomakha

***Abstract:*** Imaging of the human body at ultra-high fields (static magnetic field $B_0 \geq 7$ Tesla) is challenging due to the radio-frequency field inhomogeneities in the human body tissues caused by the short wavelength. These effects could be partially mitigated using an array of antennas and by parallel transmission allowing for control of the radio-frequency field distribution in the region of interest. All commonly-used radio-frequency arrays for ultra-high field MRI consist of resonant elements: dipoles, TEM-resonators, loops and individual slots. All these elements rely on excitation in the reactive near-field region, in the sense that they are resonant devices that produce a field pattern with a fixed relative phase distribution between the currents in different parts of the structure along the commensurable conductor elements. In our work, we propose to use a previously demonstrated non-resonant leaky-wave approach to control the phase of currents in radio-frequency coil conductors to increase the $B_1^+$ field in the center or in the region of interest. Using this approach, we developed a radio-frequency coil based on a leaky-wave antenna approach with optimized surface current distribution resulting in stronger $B_1^+$ in the desired region compared to e.g. a fractionated dipole.

## I. Introduction

Magnetic resonance imaging (MRI) is a rapidly developing, non-invasive technology for diagnosis of various diseases as well as biomedical investigation of the human body. One of the current trends is increasing the static field $B_0$ of superconductive magnets, which can provide a higher signal-to-noise ratio (SNR) to achieve higher resolution, as well as improved tissue contrast [1]. Ultra-high-field (UHF) MRI with a static magnetic field $B_0$ of 7T has become available for clinical diagnostics of the human head and extremities [2]. UHF MRI has been successfully employed for the investigation of the human brain [3] and early detection of cancer [4]. However, challenges remain in the development of radio-frequency (RF) hardware, and RF safety concerns have slowed down the application of UHF MRI to some clinical tasks, especially for whole body imaging.

K. Popova is the independent Researcher (e-mail: kristina.shin@metalab.ifmo.ru).

R. Balafendiev is the independent Researcher (e-mail: rub8@hi.is).

J. T. Svejda is with General and Theoretical Electrical Engineering (ATE), Faculty of Engineering, University of Duisburg-Essen, and Center for Nanointegration Duisburg-Essen (CENIDE), 47048 Duisburg, Germany (e-mail: jan.svejda@uni-due.de).

A. Rennings is with General and Theoretical Electrical Engineering (ATE), Faculty of Engineering, University of Duisburg-Essen, and Center for Nanointegration Duisburg-Essen (CENIDE), 47048 Duisburg, Germany (e-mail: andre.rennings@uni-duisburg-essen.de).

A. J. Raaijmakers is with Imaging Division, UMC Utrecht, Utrecht, the Netherlands, and also with Medical Image Analysis, Biomedical Engineering, Technical University of Eindhoven, Eindhoven, The Netherlands (e-mail: A.J.E.Raaijmakers@tue.nl).

C. M. Collins is with Center for Advanced Imaging Innovation and Research (CAI2R) and the Bernard and Irene Schwartz Center for Biomedical Imaging, Department of Radiology, New York University Grossman School of Medicine, New York, NY, USA (e-mail: c.collins@nyumc.org).

R. Lattanzi is with Center for Advanced Imaging Innovation and Research (CAI2R) and the Bernard and Irene Schwartz Center for Biomedical Imaging, Department of Radiology, New York University Grossman School of Medicine, New York, NY, USA (e-mail: Riccardo.lattanzi@nyumc.org).

D. Erni is with General and Theoretical Electrical Engineering (ATE), Faculty of Engineering, University of Duisburg-Essen, and Center for Nanointegration Duisburg-Essen (CENIDE), 47048 Duisburg, Germany (e-mail: daniel.erni@uni-due.de).

G. Solomakha is with High-field Magnetic Resonance, Max Planck Institute for Biological Cybernetics, Tuebingen, Germany (e-mail: georgiy.solomakha@tuebingen.mpg.de).

Classical volumetric RF coils used for whole body imaging are driven by only two independent channels (or only one separated into quadrature) to generate a circularly polarized (CP) field in the human body. However, previous work [5] showed that volumetric quadrature transmit (Tx) RF coils can create a strongly inhomogeneous field distribution in the human body at 7T. To address the inhomogeneity of the transmit $B_1^+$ field, a parallel transmission (pTx) approach was introduced [6] using arrays of individually driven transmit RF coils located at the surface of the human body. In pTx, phase and amplitude of the individual channel excitations can be optimized together in order to homogenize the $B_1^+$ field in a desired region of interest (ROI) or to minimize specific absorption rate (*SAR*) [7].

Different RF coil elements can be used for pTx arrays, including loops [8], stripline resonators [9], dipoles [10] and slot antennas [11]. Individual array elements are usually optimized to have maximum $B_1^+$ in the target imaging region and minimal *SAR* at the region located close to the surface during transmission. It has been shown [10] that dipole antennas are more efficient for the imaging of deeply located regions such as the prostate and the heart at UHF in comparison to the stripline and loop RF coils because of their less concentrated current distribution generated at the surface of the subject. Several configurations of dipole antennas for 7T body imaging were introduced during the last years, including fractionated [12], dual-mode [13] and passive-fed [14].

In UHF MRI, the ideal RF surface current patterns that maximize performance in the center of a tissue-mimicking sample tend to have currents that lead to a gradual phase evolution along the Z-axis of MR-magnet [15], [16]. This lead to an idea, that to create element, that will combine signals with optimal phases (in receive) or will focus field in desired region (in transmit) also should have non-uniform phase along z-axis. However, all above-mentioned designs are based on the resonant current distributions that occur in the conductor, generating standing waves that lead to a constant phase of the surface current.

To overcome this issue, a non-resonant leaky wave antenna (LWA) was successfully proposed as a versatile RF coil element [17]. Such an antenna realizes a linear phase distribution of the surface current due to its intrinsic non-resonant behavior, has a broad impedance matching bandwidth, and is easy to match [17]. In this work we propose a novel RF coil based on LWA for UHF body imaging using a non-uniform phase in the current distribution with maximized $B_1^+$ field in the ROI during transmission.

## II. Methods

### A. Optimizing current patterns at ultra-high-field MRI

Ultimate performance limits, such as the UISNR [15], [18]-[23], the ultimate intrinsic specific absorption rate [7], [24] and the optimal transmit efficiency (OptTx) [16] have been proposed as absolute metrics to assess optimal RF coil designs. To calculate these performance benchmarks, an optimization process is performed that involves combining electromagnetic fields generated by basis modes. These modes can be represented by fundamental surface current patterns, which are treated as associated to individual coil elements of a hypothetical infinite array. To mimic body imaging, the current basis can be defined on a cylindrical surface surrounding a homogenous tissue-mimicking cylindrical object (i.e. a so-called phantom). The optimal weighted combination of the basis modes that yields the best possible performance for a task of interest defines the ideal surface current distribution. Inspired by the approach of ideal current patterns (ICPs) originally developed for UISNR [15], [25], we generated a current pattern in the transmit case (Optimal current pattern, OCP), where the goal is to maximize the $B_1^+$ field in the central region of the sample while minimizing the total absorbed power for a fixed accepted power.

As it was mentioned in the introduction, in UHF MRI, the OCP, namely the RF surface current patterns that maximize performance (i.e., the level of $B_1^+$ field) at the center of the phantom tend to have currents that lead in phase towards the ends and lag in phase towards the middle of the current pattern. This could be explained from a geometrical point of view: if we consider several discrete current samples arranged along the bore axis ($z$-axis) at the operation frequency of a 7T MRI (298 MHz), their phase increases from the center toward the ends of a cylindrical phantom to provide focusing of the $B_1^+$ field in the object's center.

In contrast, resonant MRI coils and array elements commonly used for RF magnetic field excitation and spin relaxation signal reception (i.e. birdcage RF coils, arrays of dipoles, loops and stripline antennas) have current distributions with constant phase in the longitudinal direction [26-28]. One prior attempt to approximate the ICP was to use curved dipoles placed on top of a curved dielectric substrate so that the central portion would be further from the sample than the ends [29]. This aims at a geometric approach to change the current phase distribution along the $z$-direction. If we imagine such a dipole antenna as a combination of multiple discrete radiators, we can describe the effect of increased field as a result of constructive interference at the phantom's center. However, such an approach requires significant space inside the MR system compared to the traditional dipole array, and, therefore, is not practical.

In reality, it is quite difficult to realize these ICP, predicted by ideal-case calculations. According to one work [25] for a large cylindrical phantom that mimics the human body, Tx arrays of 8 and 16 loops are reaching only 45% and 62% of the maximal possible UISNR at the center. Arrays based on straight dipole antennas with the same lengths as loops (22 cm) are more feasible for reaching UISNR, with 8 dipoles achieving 75% of UISNR and 16 dipoles accordingly 81% of UISNR at the center. To approach OptTx, several methods could be utilized. One possible method is using double-row arrays of dipoles with complicated decoupling circuits [30]. However, this is problematic since the majority of modern UHF MRI systems have only eight available Tx-channels.

In our prior work [17], a non-resonant approach for RF field excitation was proposed using a LWA. The LWA is based on the excitation of an array of slots, which are fed by a travelling wave on a microstrip line (MSL). In this LWA coil, there is no standing wave present. The phase delay between slots is defined by the electrical length of the intervening MSL sections. In [17], a linear phase distribution was used. In this paper, a similar approach was taken for the RF coil design, but with an optimized, non-uniform phase distribution to realize an optimal current distribution. We do not aim to fully mimic ICPs, but rather introduce a platform based on a LWA design that offers the versatility to realize different non-uniform current distributions within a single RF coil element by controlling the excitation phases of its individual slot elements using a series feeding network. Phase of slot excitation can be controlled by changing the length of phase shifters between slots. For demonstration, we focused our study on the optimization of the RF magnetic field in transmission by generating specific OCP.

### *B. Simulations of multi-source RF coil with optimized current distribution*

#### *i. Multi-port LWA structure and feeding/termination*

As a starting point, we modeled a LWA based on previous work [31]. We designed an array consisting of six individually-driven, etched I-shaped slot antennas, located 2 cm above the surface of a cylindrical homogeneous phantom (ε=60, σ=0.6 S/m), mimicking the human body at 298 MHz. The gap between the I-shaped slot antennas and the phantom surface was chosen to accommodate the physical thickness of the future coil prototype and the phantom wall, corresponding to the practical constraints. I-shaped slots were used to increase the real part of the individual slot's input impedance and to improve the loading of the RF coil onto the phantom. The slots were created by removing material from the shared metal ground plane and each I-shaped slot was designed with an overall length of L=115 mm. The slots were spaced with a 60 mm period and distributed uniformly along the *z*-axis. Each slot in the implemented design was driven using a lumped port, meaning that the slots were considered as individually driven antennas. Fig. 1 illustrates the configuration of the proposed RF coil.

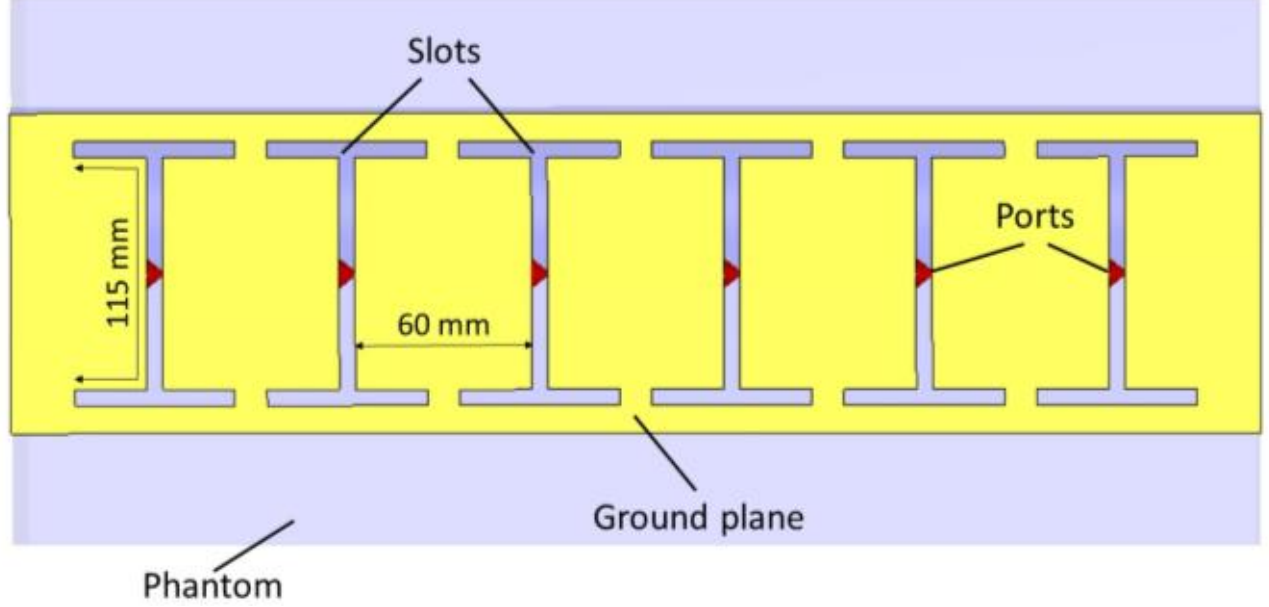

Fig. 1. General view of the proposed RF coil placed over a homogeneous phantom.

The numerical simulations for the slot antenna were performed with CST Studio Suite 2020 (Dassault Systèmes, Vélizy-Villacoublay, France) using the frequency domain solver. Initially, each slot of the antenna was driven individually to obtain the RF field created by each port.

#### *ii. Optimization of surface current distribution*

The ROI was set to a point at a depth of 7 cm to correspond to the ROI in prostate imaging applications. In order to maximize the ratio of the RF field in the ROI to the accepted power (i.e., to maximize Tx efficiency) we exported the field distributions resulting from individual slot excitation in tandem with the S-matrix of the model for subsequent use in the optimization process. Using the Python 3.7 SciPy library implementation of the Trust-Region Constrained algorithm, we solved the minimization problem for six variables, namely the amplitude and the phase input coefficients, which are the initial values for the excitation of each slot, for three slots (owing to the symmetry of the problem, the ROI was assumed to be located directly under the center of the coil). The six variables can then be further reduced to four since both the amplitudes and the phases of two of the three slots are defined relative to the centermost one. Within the target function, the fields of the individual slots were first combined with amplitude and phase coefficients to find the total RF field value in the ROI and then the same input coefficients were multiplied by the S-parameter matrix to find the output coefficients, which represent the adjusted amplitudes and phases of the excitation signals for each slot after accounting for the mutual coupling between them, as determined by the S-parameter matrix, and the accepted power.

The optimized surface current distribution was obtained by recalculating the CST model of the coil with the optimized input coefficients (from edge to center: 0.68, 0.927 and 1 √W in amplitude and 42.1°, 27.1° and 0° in phase) and extracting the surface current along the surface of the phantom. The resulting 1D current distribution $J_z(z)$ was then imported into a corresponding COMSOL Multiphysics (Burlington, MA, USA) model to estimate the impact of current discretization introduced by the slots. This current distribution was then extended to a CP-mode producing a 2D current distribution continuous along *φ,* resulting in a current strongly resembling the ICP. We then compared the resulting optimized current, in which the phase increases towards the ends, with the case of a constant current phase (along the *z*-axis). Phases of the surface current $J_z$ at the top surface of the phantom for different cases are shown in Fig. 2. The current was discretized, first by discrete angle *φ* and then by both discrete *φ* and *z*, to estimate if the improvement introduced by the variable phase persists when the field in the phantom is induced by short sections of the current, as is the case in a slot antenna. In general, generated OCP is looking very similar to ICP for reaching UISNR in the center of this cylindrical phantom [25].

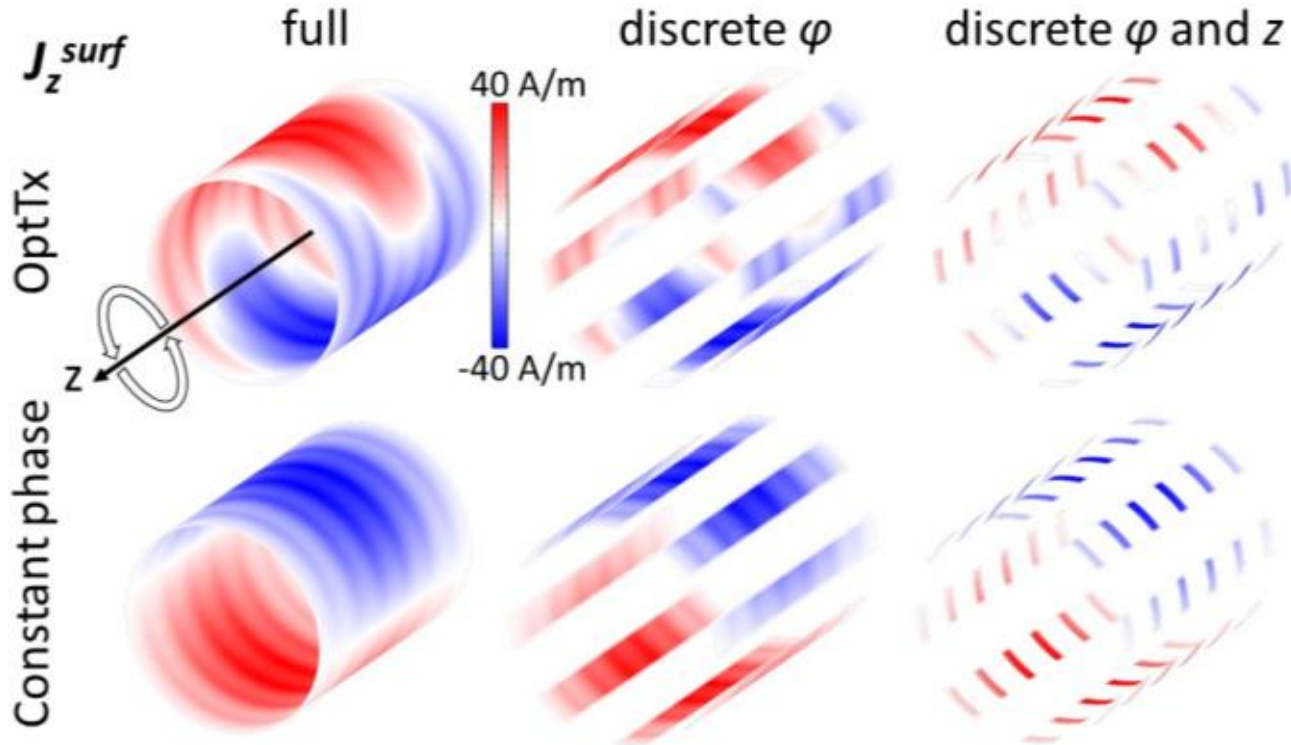

Fig. 2. Current distributions used in COMSOL simulations. The columns represent different current discretization schemes: continuous full distribution, discretization by *φ*, and discretization by both *φ* and the *z*. The rows show different phase conditions, namely an optimized current with phase increasing towards the ends, and a constant phase current along the *z*-axis. The color scale represents the magnitude and direction of the current density, with red indicating positive values and blue indicating negative values.

Fig. 3 shows the distribution of the RF magnetic field $B_1^+$ (normalized by the square root of the accepted power $P_{acc}$ calculated as the integrated power loss density in the phantom) simulated for the current configurations in Fig. 2.

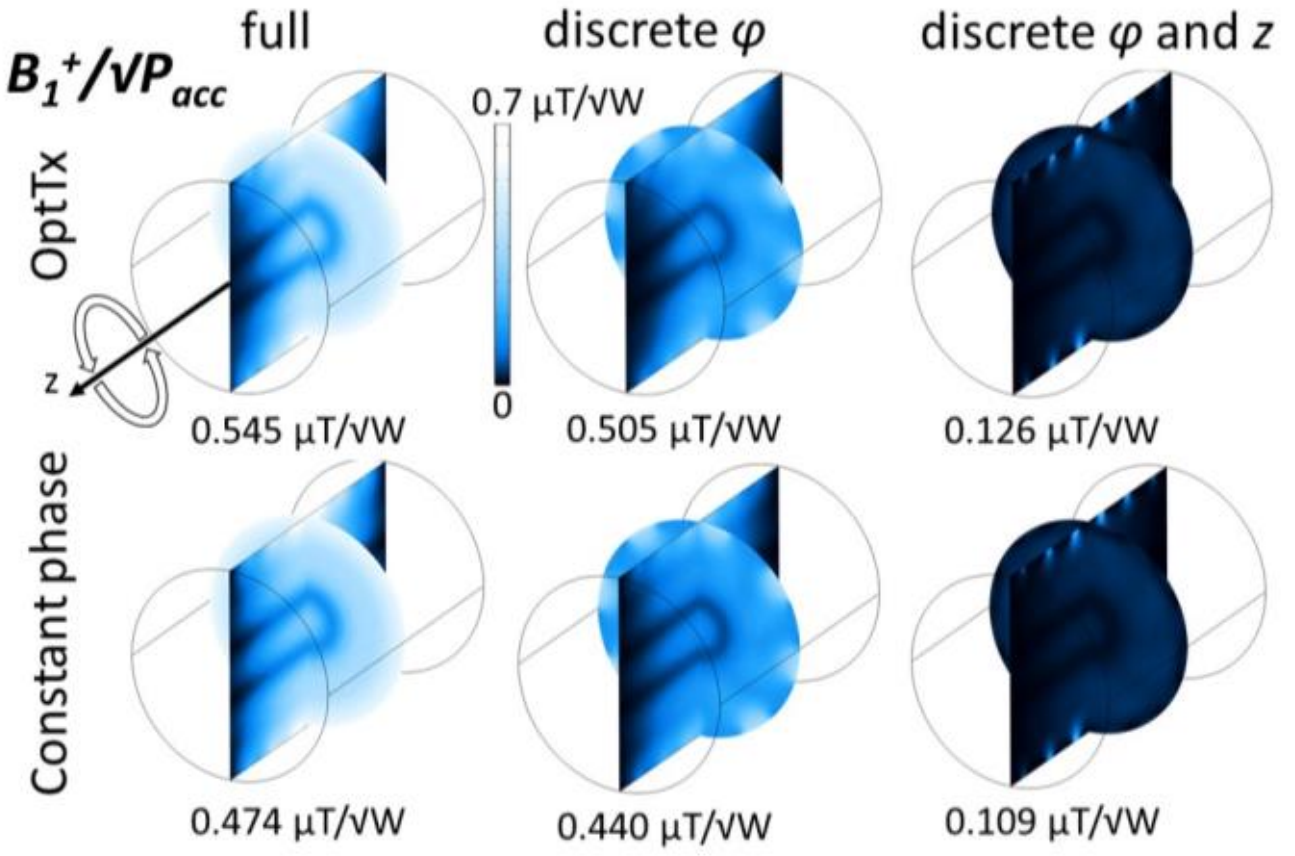

Fig. 3. Simulated $B_1^+/\sqrt{P_{acc}}$ for the current distributions. The value at the center of the phantom is listed below each case.

Each image presents the normalized magnetic field $B_1^+$ with a variable (optimal) and constant current phase along the $z$-axis. The normalized magnetic field $B_1^+$ for the full case is evenly distributed in the simulation region, with smooth intensity changes. For the full current distribution, the OptTx $B_1^+$ at the center is 15% larger than that for the case with constant phase. When discretizing by $\varphi$, the OptTx $B_1^+$ is about 14.8% larger than that of the constant-phase case. With discretization along both $\varphi$ and $z$, the OptTx $B_1^+$ field amplitude significantly decreases. Nonetheless, even with two-parameter discretization with a variable phase, the OptTx $B_1^+$ is still 15.6% larger than for the current distribution with constant phase. Thus, independently from the current discretization, having the phase that increases towards the ends of the antenna has an advantage in terms of the OptTx $B_1^+$ field amplitude.

### *C. Design of single-source RF coil with optimal currents*

To demonstrate that the LWA can be used to create an optimal current distribution at 298 MHz, based on numerical simulation results, we designed an OptTx-coil as a MSL section with a periodically slotted ground plane. The name *OptTx* was chosen since the proposed coil was designed to approach the optimal Tx-efficiency in a region of interest (at a depth of 7 cm). The final version of the proposed OptTx-coil is shown in Fig. 4.

We designed the OptTx-coil inspired by previous work [31] for an octagonal (40 cm × 20 cm × 40 cm) pelvis-shaped phantom (ε=60, σ=0.6 S/m) that mimics the human pelvis [11]. We used a length of 400 mm as in [31], because it is the maximum possible length along which the exponentially decaying wave propagates along the MSL to maximize power radiation before the terminal load. The length and spacing of the slots etched in the ground plane were adjusted to optimize efficiency in the desired ROI at a depth of 7 cm, resulting in a width of 145 mm and an overall length of 390 mm. The LWA coil element features a ground plane with six identical I-shaped slots, as in the numerical model. These slots are used to increase the power coupling efficiency per unit length. To mitigate *SAR* and isolate the LWA coil's slotted ground plane from the phantom, a 15-mm polycarbonate spacer was introduced.

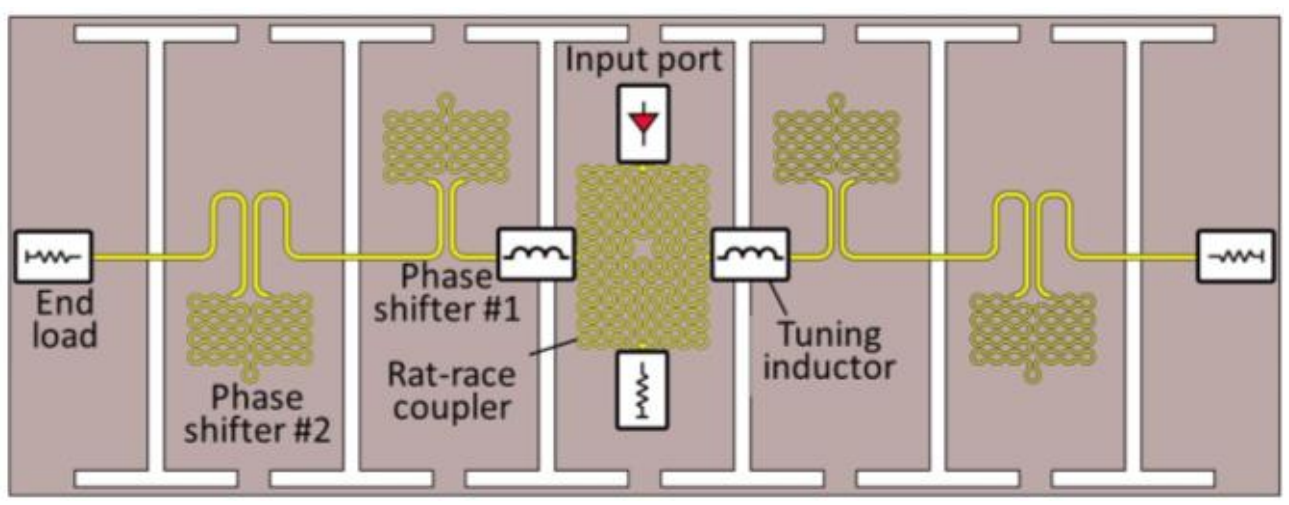

Fig. 4. Proposed OptTx-coil with optimized current phase design.

Note that a 145 mm width of the OptTx-coil ensures efficient coupling with the load since the proposed LWA is effectively two times shorter than previously reported LWA [31] to realize symmetrical current distribution. This width, however, can limit the number of elements in an array configuration. Future designs can be miniaturized by using higher-permittivity spacer materials, allowing for reduced coil dimensions without compromising $B_1^+$ performance.

In the proposed OptTx-coil, the guiding transmission line with a strip height of 2 mm was designed to achieve a 50 Ω impedance to match the output impedance of the MRI-scanner transceiver system. It was separated from the transmission line by a 2-mm-thick foam layer. Since it is necessary to provide a growing phase along the axial direction from the center of the OptTx-coil towards its end, we feed two sections of a proposed OptTx-coil from the center with a 180° phase shift. To realize the signal splitting with the required phase shift, an ultra-compact rat-race coupler was used as proposed previously in [32]. From the output ports of the rat-race coupler, the signal is led to the sections of the OptTx-coil. Additional lumped inductors were placed at the inputs of the leaky-wave sections to provide impedance matching. The holes passing through both sides of the OptTx-coil are used to fasten the top and bottom PCBs using dielectric screws.

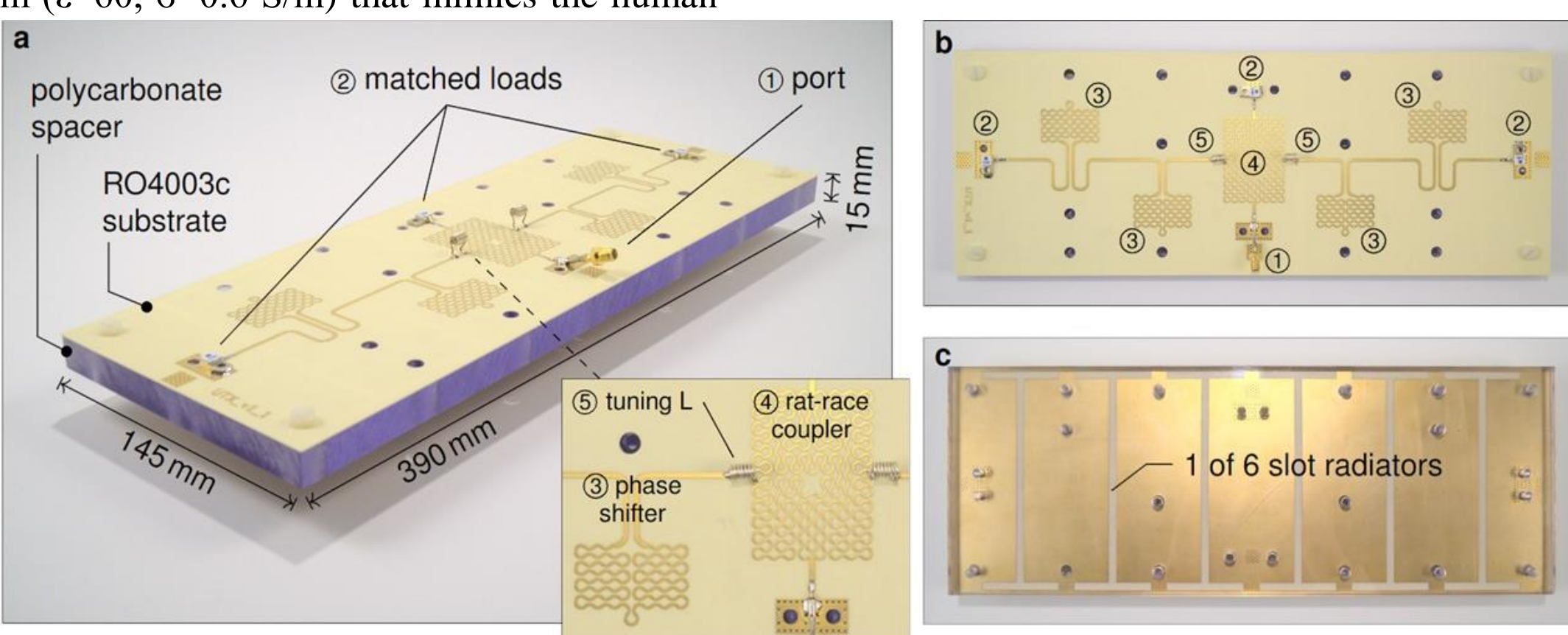

Fig. 5. Photo of the prototype of OptTx-coil in (a) a perspective view, (b) a top view, and (c) a bottom view.

A similar approach as in [31] regarding the transmit field optimization was used: single slot $B_1^+$ field profiles were imported to a Python 3.7 script and then optimized to maximize the $B_1^+$ field in the region, located at 7 cm depth beneath the antenna center. To provide the desired phase shift, which was obtained by the optimization process, microstrip phase shifters were used. Since the phase lag from the center of the OptTx-coil towards its end should be positive, phase shifters with values larger than 180° were used. Phase shifters were implemented as a combination of a 180° 70.7 Ω line and a 50 Ω line. The 70.7 Ω line was used due to its better miniaturization ability because of its smaller MSL width.

### D. RF coil prototyping and bench measurements

To construct the OptTx-coil, two 0.2 mm thick Rogers RO4003 low-loss dielectric PCBs (ε=3.38) were positioned on top of a 2-mm-thick layer of ROHACELL 31HF foam. The top PCB was responsible for forming the stripline, which included the ultra-compact rat-race coupler, the lumped inductors, and the microstrip phase shifters. The bottom PCB served as the ground plane and featured six etched I-shaped slots. A 15-mm-thick polycarbonate spacer was inserted to separate the slotted ground plane of the OptTx-coil from the phantom, helping to control the *SAR* levels. The overall design of the OptTx-coil prototype is displayed in Fig. 5.

The LWA was powered from one side using a 50 Ω coaxial cable, soldered to the MSL. The remaining three ports of the LWA were terminated with 50 Ω high-power resistors. They were placed at the inputs of LWA sections to provide impedance matching. No matching network or balun was required in this prototype.

For all tests, a homogeneous phantom with dimensions and electrical properties similar to those used in the numerical simulations was utilized. To experimentally test the performance of the proposed OptTx-coil and the reference fractionated dipole, the RF coils and the phantom were placed in a Philips Achieva 7T human MRI scanner. Single channels were used for transmission and reception. The $B_1^+$ maps were obtained using the Dual-TR actual flip-angle imaging (AFI) method. The acquisition parameters were as follows: a FOV = (231×400×140) mm$^3$, a voxel size of (2.2×3.8×10) mm$^3$, an TE = 2.2 μs, TR1 = 50 μs, TR2 = 250 μs, with a total forward power of 79 W.

## III. Results

The optimized OptTx-coil element was numerically and experimentally compared to a fractionated dipole using the aforementioned homogeneous phantom within the 7T MRI scanner. The amplitude of the RF $B_1^+$ field was normalized by the square root of the accepted power in the central coronal slice plane at the center of the phantom and is presented in Fig. 6a and 6b for both the fractionated dipole and the proposed OptTx-coil with two optimization methods applied.

For analyzing the RF field distribution at different depths, we extracted RF $B_1^+$ field profiles through the centers of the ROI. Fig. 6c presents these profiles, comparing the OptTx-coil and the fractionated dipole (dashed lines in Fig. 6a and 6b indicate the distance). Compared to the fractionated dipole, the OptTx-coil generates a RF $B_1^+$ field with higher magnitude at every depth, including the ROI at a 7-cm depth.

The simulated and measured $B_1^+$ amplitude for the proposed OptTx-coil and the fractionated dipole are compared in Table 1. All values correspond to the $B_1^+$ amplitude at a single point located at the center of the phantom at a depth of 7 cm, normalized to 1 W accepted power to enable direct comparison.

TABLE I
COMPARISON BETWEEN SIMULATED AND MEASURED $B_1^+$ AMPLITUDE FOR THE PROPOSED OPTTX-COIL AND THE FRACTIONATED DIPOLE

| RF coil type | Simulated $\|B_1^+\|$, μT | Measured $\|B_1^+\|$, μT |
|---|---|---|
| OptTx | 0.315 | 0.335 |
| Dipole | 0.249 | 0.285 |

This table shows that the RF field generated by the OptTx-coil at a depth of 7 cm (corresponding to the prostate) was 26.5% and 17.5% stronger than that of the fractionated dipole (in case of the numerical modeling and of the experiment, respectively).

We evaluated $pSAR_{10g}$ (using the CST Legacy method,

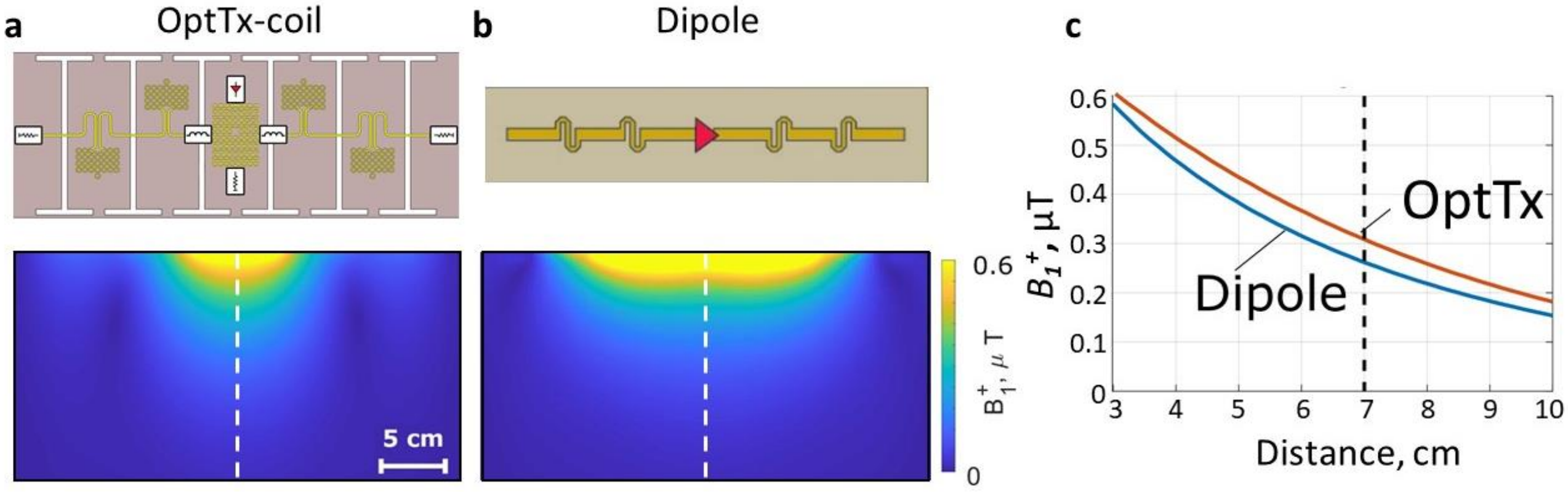

Fig. 6. Simulated $B_1^+$ field distribution for (a) the proposed OptTx-coil element and (b) the fractionated dipole, and (c) the associated $B_1^+$ field profiles at 7-cm depth.

normalized to 1 W accepted power) and SAR-efficiency (mean $B_1^+$ normalized to $\sqrt{\text{pSAR}_{10g}}$) for both coils with a homogeneous phantom. The $\text{pSAR}_{10g}$ values were 1.276 W/kg for the OptTx-coil and 1.244 W/kg for the dipole. SAR-efficiency at 7 cm depth reached 0.279 μT/√(W/kg) for the OptTx-coil and 0.223 μT/√(W/kg) for the dipole, corresponding to a 25.1% improvement.

In our previous work [31], we showed the difference between the uniform phase distribution of the dipole and a non-uniform distribution of the slot array. Fig. 7 depicts the RF $B_1^+$ field for the non-uniform phase distribution of the OptTx-coil element showing a narrower field of view but with significantly larger magnitude at large depths in comparison to the reference dipole for the same $P_{acc}$.

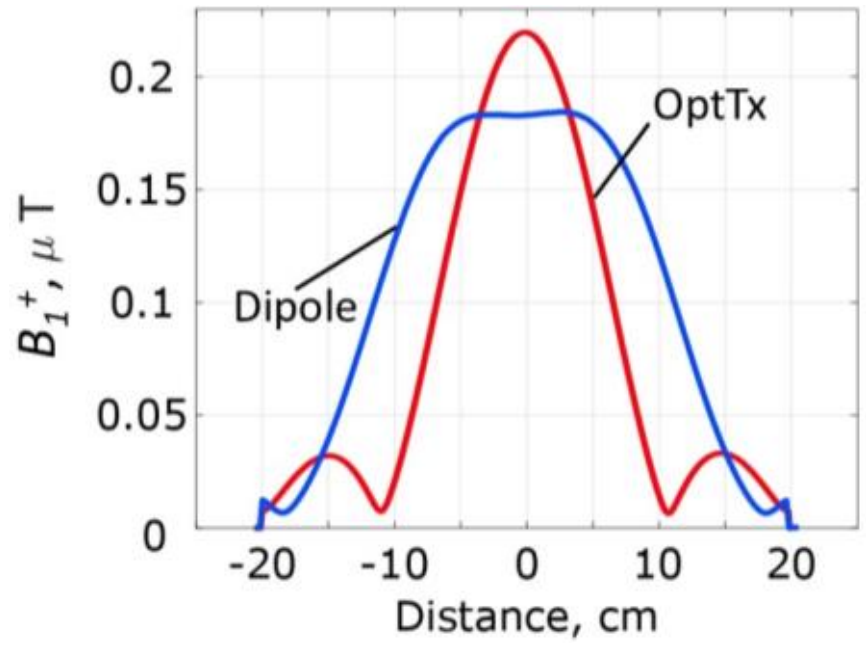

Fig. 7. $B_1^+$ amplitude for the proposed OptTx-coil and for the fractionated dipole along the *z*-axis at 7-cm depth.

The experimentally measured distributions of the RF $B_1^+$ field in the central cross-section of the phantom for the OptTx-coil element and the fractionated dipole are presented in Fig. 8a and 8b, respectively (the phantom is located in the entire region of the $B_1^+$ map). Fig. 8c shows depth profiles both for the proposed OptTx and dipole coils (dashed lines in Fig. 8a and 8b indicate the distance).

evaluated. Comparing the optimized phase distribution with a constant phase distribution as visualized in Fig. 2, confirmed the advantages of the optimized configuration in enhancing the RF $B_1^+$ field strength within the target region.

We found that the optimized phase of the OptTx-coil progressively increases from the center to the edges. The optimization of current using a variable phase resulted in an approximately 15% increase in Tx-efficiency compared to the case with constant phase along the *z*-axis. This indicates that by increasing the phase of current towards the ends of the RF coil, more magnitude of the $B_1^+$ field can be focused on the ROI.

In comparison, the fractionated dipole provides a more uniform magnetic field distribution along its length, but the dipole is less effective at deeper penetration in the ROI. Experimental results further shown in Fig. 8 demonstrate that the OptTx-coil creates an RF $B_1^+$ field with higher amplitude at all depths, including the ROI, by 17.5% compared to the dipole (at 7 cm depth), confirming and even surpassing the numerical simulation results. Some discrepancies between simulation and experiment might be caused by possible differences in phantom dielectric properties, minor positioning inaccuracies, non-ideal matching and additional losses in the dipole's feeding network, and inaccuracies in the experimental settings of $B_1^+$ mapping, which was optimized for a specific 7 cm depth.

The OptTx-coil is particularly effective for MRI of an ROI at significant tissue depth, such as the prostate, due to its ability to concentrate the RF $B_1^+$ field in the ROI and its improved directionality. This OptTx-coil design maximizes the use of the wavelength to focused RF $B_1^+$ field with higher intensity in the ROI before the wave fully decays.

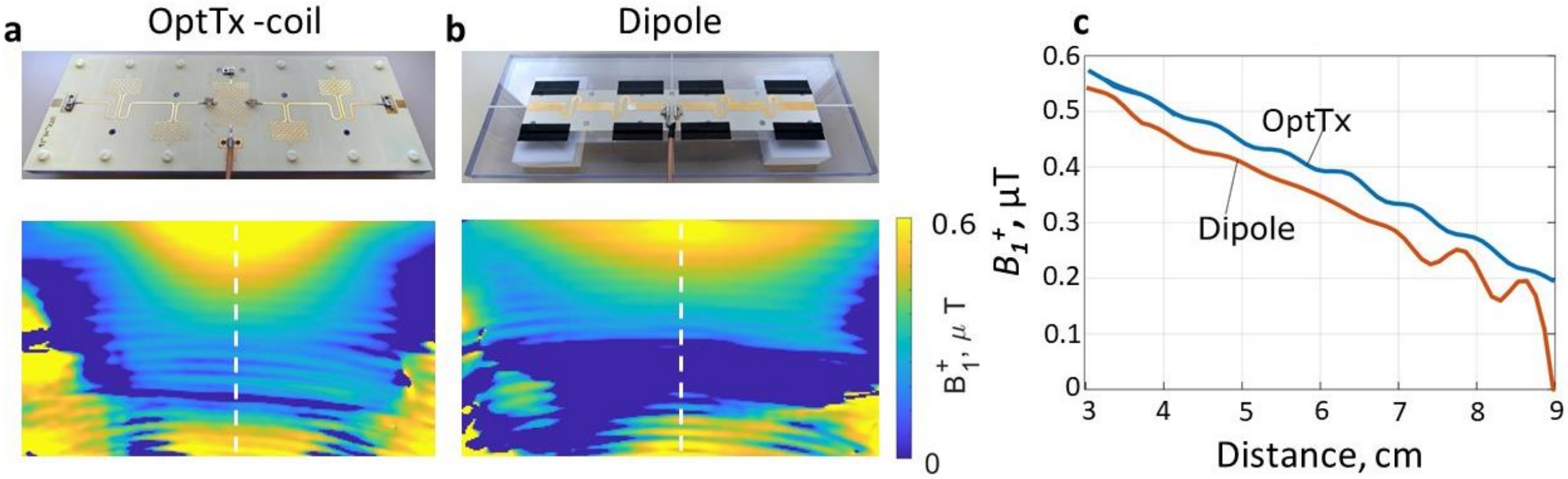

Fig. 8. Measured $B_1^+$ field distribution for (a) the proposed OptTx-coil element and (b) the fractionated dipole, and (c) the associated $B_1^+$ field profiles at 7-cm depth.

## IV. Discussion and Conclusion

To maximize the transmit $B_1^+$ field in the ROI located at a depth of 7 cm into the phantom (relevant to prostate imaging), we employed a multistep approach. First, the individual $B_1^+$ field distributions and S-matrix parameters of the six-slot antenna array were processed using the Trust-Region Constrained algorithm to determine the OCP for maximizing Tx-efficiency. Subsequently, the impact of discretizing this continuous optimal current distribution on the RF $B_1^+$ field characteristics and Tx-efficiency was

However, by focusing on the Tx-field, we are limiting longitudinal coverage. This limits proposed approach to application with relatively small longitudinal FOV. However, in general, UHF body imaging is limited to small volume applications such as prostate, heart and liver imaging [10].

The OptTx-coil is specifically engineered to have a phase increase from the center to the ends, which enhances the transmitted signal's intensity. The OptTx-coil design provides a unique approach for MRI. The use of an ultra-compact rat-race coupler, based on the work in [32], in conjunction with microstrip phase shifters, enables the generation of a tailored phase distribution that maximizes power radiation and minimizes *SAR*. However, the inclusion of phase shifters and the miniaturized rat-race coupler, which ensure the necessary phase shifts and improve impedance matching with the MRI system, significantly complicates the OptTx-coil design.

In future work, we intend to create an 8-channel transceiver body RF coil based on this LWA approach, to improve the magnitude of the RF field in a desired ROI such as the prostate.

## Acknowledgment

The authors also gratefully acknowledge Dr. Carel van Leuween and Prof. Bart Steensma for their help in setting up MR-measurements and data post-processing, and Prof. Stanislav Glybovski for his help in theoretical studies and valuable discussions.

## References

[1] M. E. Ladd, P. Bachert, M. Meyerspeer, E. Moser, A. M. Nagel, D. G. Norris, S. Schmitter, O. Speck, S. Straub, and M. Zaiss, "Pros and cons of ultra-high-field MRI/MRS for human application," Progress in Nuclear Magnetic Resonance Spectroscopy, vol. 109, pp. 1–50, 2018.

[2] G. A. Keith, R. A. Woodward, T. Hopkins, S. Allwood-Spiers, J. Trinder, K. W. Muir, D. A. Porter, N. E. Fullerton, "Towards clinical translation of 7 Tesla MRI in the human brain, IPEM-Translation," vol. 9, 2024.

[3] A. Abosch, E. Yacoub, K. Ugurbil, and N. Harel, "An assessment of current brain targets for deep brain stimulation surgery with susceptibility-weighted imaging at 7 Tesla," Neurosurgery, vol. 67, no. 6, pp. 1745–1756, 2010.

[4] S. Gruber, K. Pinker, O. Zaric, L. Minarikova, M. Chmelik, P. Baltzer, R. N. Boubela, T. Helbich, W. Bogner, and S. Trattnig, "Dynamic contrast-enhanced magnetic resonance imaging of breast tumors at 3 and 7 T: a comparison," Investigative Radiology, vol. 49, no. 5, pp. 354–362, 2014.

[5] J. T. Vaughan, C. J. Snyder, L. J. DelaBarre, P. J. Bolan, J. Tian, L. Bolinger, G. Adriany, P. Andersen, J. Strupp, and K. Ugurbil, "Whole-body imaging at 7T: Preliminary results," Magn. Reson. Med., vol. 61, no. 1, pp. 244–248, 2009.

[6] Y. Zhu, "Parallel excitation with an array of transmit coils," Magn. Reson. Med., vol. 51, no. 4, pp. 775–784, 2004.

[7] R. Lattanzi, D. K. Sodickson, A. K. Grant, and Y. Zhu, "Electrodynamic constraints on homogeneity and radiofrequency power deposition in multiple coil excitations," Magn. Reson. Med., vol. 61, no. 2, pp. 315–334, 2009.

[8] R. G. Pinkerton, E. A. Barberi, and R. S. Menon, "Transceive surface coil array for magnetic resonance imaging of the human brain at 4 T," Magn. Reson. Med., vol. 54, no. 2, pp. 499–503, 2005.

[9] G. Adriany, P.-F. Van de Moortele, F. Wiesinger, S. Moeller, J. P. Strupp, P. Andersen, C. Snyder, X. Zhang, W. Chen, K. P. Pruessmann et al., "Transmit and receive transmission line arrays for 7 Tesla parallel imaging," Magn. Reson. Med., vol. 53, no. 2, pp. 434–445, 2005.

[10] A. Raaijmakers, O. Ipek, D. Klomp, C. Possanzini, P. Harvey, J. Lagendijk, and C. Van den Berg, "Design of a radiative surface coil array element at 7 T: the single-side adapted dipole antenna," Magn. Reson. Med., vol. 66, no. 5, pp. 1488–1497, 2011.

[11] L. Alon, R. Lattanzi, K. Lakshmanan, R. Brown, C. M. Deniz, D. K. Sodickson, and C. M. Collins, "Transverse slot antennas for high field MRI," Magn. Reson. Med., vol. 80, no. 3, pp. 1233–1242, 2018.

[12] A. J. Raaijmakers, M. Italiaander, I. J. Voogt, P. R. Luijten, J. M. Hoogduin, D. W. Klomp, and C. A. van den Berg, "The fractionated dipole antenna: A new antenna for body imaging at 7 Tesla," Magn. Reson. Med., vol. 75, no. 3, pp. 1366–1374, 2016.

[13] G. Solomakha, C. v. Leeuwen, A. Raaijmakers, C. Simovski, A. Popugaev, R. Abdeddaim, I. Melchakova, and S. Glybovski, "The dual-mode dipole: A new array element for 7T body imaging with reduced SAR," Magn. Reson. Med., vol. 81, no. 2, pp. 1459–1469, 2019.

[14] I. Zivkovic, C. A. de Castro, and A. Webb, "Design and characterization of an eight-element passively fed meander-dipole array with improved specific absorption rate efficiency for 7 T body imaging," NMR in Biomedicine, vol. 32, no. 8, p. e4106, 2019.

[15] R. Lattanzi and D. K. Sodickson, "Ideal current patterns yielding optimal signal-to-noise ratio and specific absorption rate in magnetic resonance imaging: computational methods and physical insights," Magn. Reson. Med., vol. 68, no. 1, pp. 286–304, 2012.

[16] I. P. Georgakis, A. G. Polimeridis, and R. Lattanzi, "A formalism to investigate the optimal transmit efficiency in radiofrequency shimming," NMR in Biomedicine, vol. 33, no. 11, p. e4383, 2020.

[17] G. Solomakha, J. T. Svejda, C. van Leeuwen, A. Rennings, A. Raaijmakers, S. Glybovski, and D. Erni, "A self-matched leaky-wave antenna for ultrahigh-field magnetic resonance imaging with low specific absorption rate," Nature Communications, vol. 12, no. 1, pp. 1–11, 2021.

[18] I. I. Giannakopoulos, I. P. Georgakis, D. K. Sodickson, and R. Lattanzi, "Computational methods for the estimation of ideal current patterns in realistic human models," Magn Reson Med, vol. 91, no. 2, pp. 760-772, 2024.

[19] M. Ohliger, A. Grant, and D. Sodickson, "Ultimate intrinsic signal-to-noise ratio for parallel MRI: electromagnetic field considerations," Magn. Reson. Med., vol. 50, no. 5, pp. 1018–1030, 2003.

[20] F. Wiesinger, P. Boesiger, and K. P. Pruessmann, "Electrodynamics and ultimate SNR in parallel MR imaging," Magn. Reson. Med., vol. 52, no. 2, pp. 376–390, 2004.

[21] B. Guerin, J. F. Villena, A. G. Polimeridis, E. Adalsteinsson, L. Daniel, J. K. White, and L. L. Wald, "The ultimate signal-to-noise ratio in realistic body models," Magn. Reson. Med., vol. 78, no. 5, pp. 1969–1980, 2017.

[22] A. Pfrommer and A. Henning, "The ultimate intrinsic signal-to-noise ratio of loop-and dipole-like current patterns in a realistic human head model," Magn. Reson. Med., vol. 80, no. 5, pp. 2122–2138, 2018.

[23] O. Ocali, E. Atalar, "Ultimate intrinsic signal-to-noise ratio in MRI," Magn. Reson. Med., vol. 30, no. 3, pp. 462-473, 1998.

[24] B. Guerin, J. F. Villena, A. G. Polimeridis, E. Adalsteinsson, L. Daniel, J. K. White, B. R. Rosen, and L. L. Wald, "Computation of ultimate SAR amplification factors for radiofrequency hyperthermia in non-uniform body models: impact of frequency and tumour location," International Journal of Hyperthermia, vol. 34, no. 1, pp. 87–100, 2018.

[25] R. Lattanzi, G. C. Wiggins, B. Zhang, Q. Duan, R. Brown, and D. K. Sodickson, "Approaching ultimate intrinsic signal-to-noise ratio with loop and dipole antennas," Magn. Reson. Med., vol. 79, no. 3, pp. 1789–1803, 2018.

[26] E. C. Hayes, W. A. Edelstein, J. F Schenck et al, "An efficient highly homogeneous radiofrequency coil for whole-body NMR imaging at 1.5 T," J Magn Reson, vol. 63, pp. 622–628, 1985.

[27] N. I. Avdievich, G. Solomakha, L. Ruhm, A. Henning, K. Scheffler, "Unshielded bent folded-end dipole 9.4 T human head transceiver array decoupled using modified passive dipoles," Magn Reson Med, vol. 86, pp. 581–597, 2021.

[28] I. Lavdas, H.C. Seton, C.R. Harrington et al, "Stripline resonator and preamplifier for preclinical magnetic resonance imaging at 4.7 T," Magn Reson Mater Phy, vol. 24, pp. 331–337, 2011.

[29] G. Chen, D. Sodickson, and G. Wiggins, "3d curved electric dipole antenna for propagation delay compensation," in Proceedings of the 23rd Annual Meeting of ISMRM, Toronto, ON, Canada, 2015.

[30] G. Chen, R. Lattanzi, D.K. Sodickson, G.C. Wiggins, "Approaching the ultimate intrinsic SNR with dense arrays of electric dipole antennas," 24th Annual Meeting of ISMRM, Singapore, 2016.

[31] G. Solomakha, R. Balafendiev, S. Glybovski, "A method for current phase manipulation in RF-Coils for UHF MRI using individually driven slots," AIP Conf. Proc. 8, vol. 2300, no. 1, 2020.

[32] A. E. Popugaev and R. Wansch, "A novel miniaturization technique in 7 microstrip feed network design," in 2009 3rd European Conference on Antennas and Propagation, 2009, pp. 2309–2313.

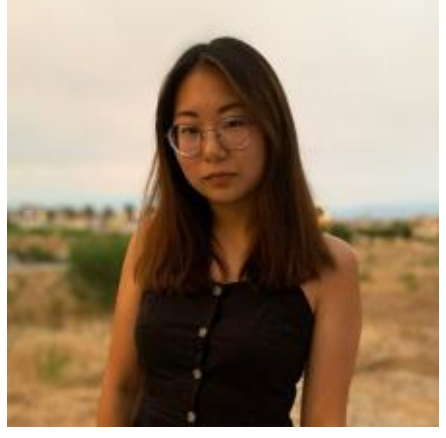

**Kristina I. Popova** was born in Saint Petersburg, Russia, in 2000. She received B.Sc. and M.Sc. degrees in technical physics from ITMO University, Saint Petersburg, Russia, in 2022 and 2024. Her research interests include the development radio frequency coils for ultra-high-field magnetic resonance imaging.

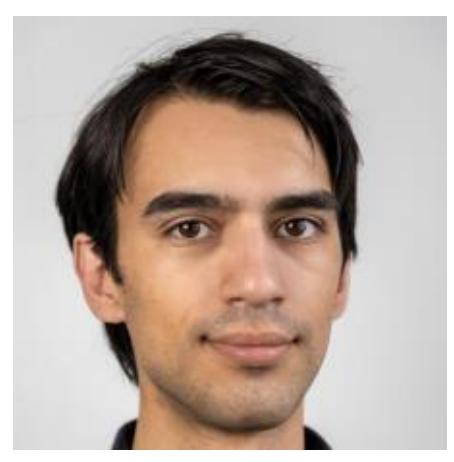

**Rustam Balafendiev** received the M.Sc. degree in radiophysics from the School of Physics and Engineering, ITMO University, Saint Petersburg, Russia, in 2021, where he is currently pursuing the Ph.D. degree. His current research interests include surface coils for ultrahigh-field magnetic resonance imaging (MRI) and novel applications of wire metamaterials.

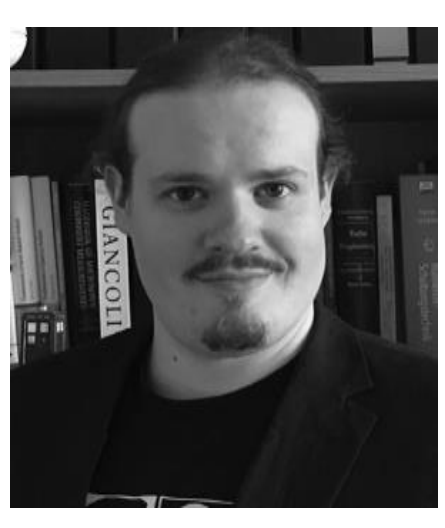

**Jan Taro Svejda** (Member, IEEE) received the B.Sc. degree in electrical engineering from the University of Applied Science, Düsseldorf, Germany, in 2008, and the M.Sc.and Dr.-Ing. degrees in electrical engineering and information technology from the University of Duisburg-Essen (UDE), Duisburg, Germany, in 2013 and 2019, respectively, for his research work in the field of X-nuclei based magnetic resonance imaging. He started his electrical engineering career with the University of Applied Science. He is currently a Research Assistant with the Department of General and Theoretical Electrical Engineering, UDE, where he is involved in teaching several lectures and courses mainly in the field of electrical engineering. His research interests include all aspects of theoretical and applied electromagnetics, currently focusing on medical applications, electromagnetic metamaterials, and scientific computing methods.

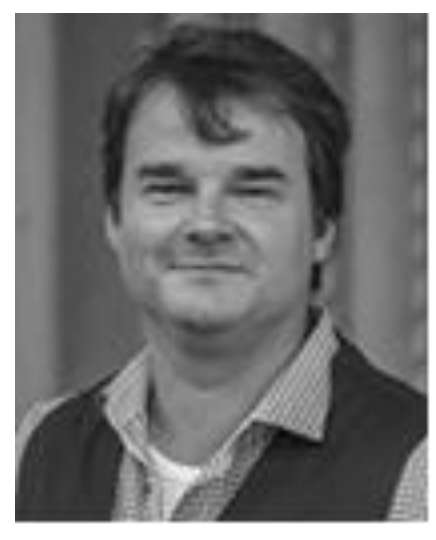

**Andreas Rennings** (Member, IEEE) received the Dipl.-Ing. and Dr.-Ing. degrees from the University of Duisburg-Essen (UDE), Duisburg, Germany, in 2000 and 2008, respectively. He studied electrical engineering with the UDE. He carried out his diploma work during a stay with the University of California in Los Angeles, Los Angeles, CA, USA. From 2006 to 2008, he was with IMST GmbH, Kamp-Lintfort, Germany, where he was an RF Engineer. Since then, he has been a Senior Scientist and Principal Investigator with the Laboratory for General and Theoretical Electrical Engineering, UDE. His research interests include all aspects of theoretical and applied electromagnetics, currently with a focus on medical applications and on-chip millimeter-wave/THz antennas. He was the recipient of several awards, including a student paper price at the 2005 IEEE Antennas and Propagation Society International Symposium and the VDE-Promotionspreis 2009 for the dissertation.

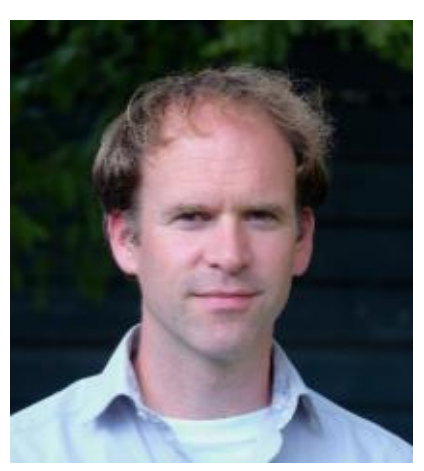

**Alexander J. E. Raaijmakers** received the M.Sc. degree in applied physics from the University of Groningen, in 2004, and the Ph.D. degree (research project) from the Radiotherapy Department, University Medical Center Utrecht. His project involved the investigation of radiotherapy dose distributions with the presence of magnetic field for (hypothetical) MRI-guided radiotherapy. He has discovered the Electron Return Effect (ERE). From 2008 to 2010, he received the Casimir Grant on the Development of New RF Transmit and Receive Coils for High-Field MRI, for which he was in part stationed with Philips Medical Systems, Best, The Netherlands. He has acquired extensive knowledge on EM physics with MRI and RF Engineering. Since 2013, he has been an Assistant Professor with the 7T Research Group, Department of Radiology, UMC Utrecht. In 2016, he was an Assistant Professor with the Research Group Medical Image Analysis, Eindhoven University of Technology. His research interest includes newly installed seven-tesla MRI scanner.

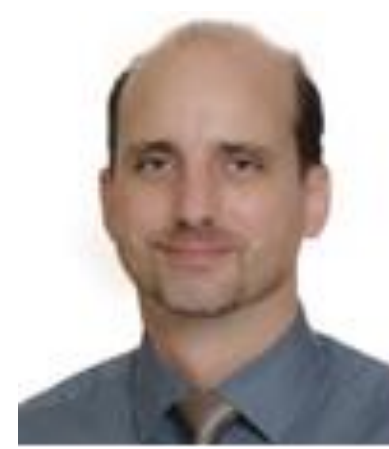

**Christopher M. Collins** received the Baccalaureate degree in engineering science from The Pennsylvania State University (PSU), University Park, USA, in 1993, and the Ph.D. degree in bioengineering from The University of Pennsylvania, Philadelphia, USA, in 1999. He joined the faculty in Radiology at PSU in 2001 after a postdoctoral fellowship there, becoming a Full Professor in 2010. He is currently a Professor of Radiology at the New York University Langone Medical Center, New York, USA, as a member of the Bernard and Irene Schwatz Center for Biomedical Imaging. His research interests include use of numerical methods considering field/tissue interactions to simulate, evaluate, and ensure safety and efficacy of MRI.

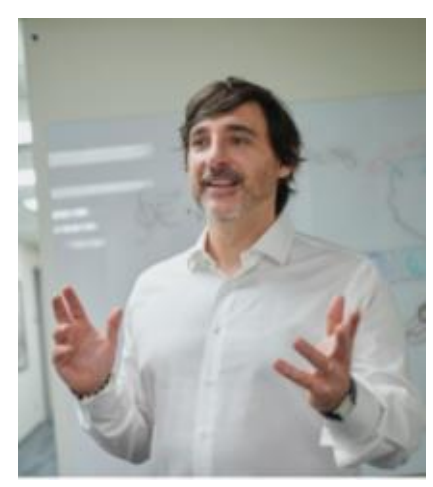

**Riccardo Lattanzi** received his laurea degree in electronic engineering from University of Bologna, his Master of Science in electrical engineering and computer science from MIT, and his Ph.D. in medical engineering and medical physics from the Harvard-MIT Division of Health Sciences and Technology. He is an Associate Professor of Radiology, Electrical and Computer Engineering at the New York University. His research work lies at the boundary between physics, engineering and medicine. He investigates fundamental principles involving the interactions of radiofrequency electromagnetic fields with biological tissue in order to develop new techniques and technologies that are aimed at improving the diagnostic power of MRI. He won the ISMRM I.I. Rabi Young Investigator Award, received an NSF CAREER award, a Fulbright scholarship, was selected as an Aspen Junior Fellow of the Aspen Institute Italia, a Young Leader of the Council for the United States and Italy and was elected to the Sigma-Xi research honor society.

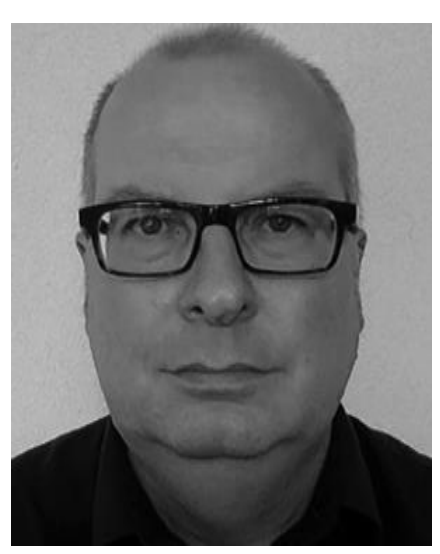

**Daniel Erni** (Member, IEEE) received the Diploma degree in electrical engineering from the University of Applied Sciences Rapperswil (OST), Rapperswil, Switzerland, in 1986, and the Diploma degree in electrical engineering and the Ph.D. degree in laser physics from ETH Zürich, Zürich, Switzerland, in 1990 and 1996, respectively. Since 1990, he has been with the Laboratory for Electromagnetic Fields and Microwave Electronics, ETH Zürich. From 1995 to 2006, he was the Founder and the Head of the Communication Photonics Group, ETH Zürich. Since October 2006, he has been a Full Professor with the Laboratory for General and Theoretical Electrical Engineering, University of Duisburg-Essen, Duisburg, Germany. Between 2017 and 2018, he joined the Institute of Electromagnetic Fields (IEF), ETH Zürich, as a Visiting Professor. He is also the Co-Founder of the spin-off company airCode, Duisburg, Germany, working on flexible printed RFID technology. His research interests include optical interconnects, nanophotonics, plasmonics, advanced solar cell concepts, optical and electromagnetic metamaterials, RF, mm-wave and THz engineering, chipless flexible RFIDs, biomedical engineering, bioelectromagnetics, marine electromagnetics, computational electromagnetics, multiscale and multiphysics modeling, numerical structural optimization, and science and technology studies. Dr. Erni is also a Fellow of the Electromagnetics Academy and a member of the Center for Nanointegration Duisburg-Essen, Materials Chain, Flagship Program of the University Alliance Ruhr, the Swiss Physical Society, German Physical Society (DPG), and Optical Society of America (Optica).

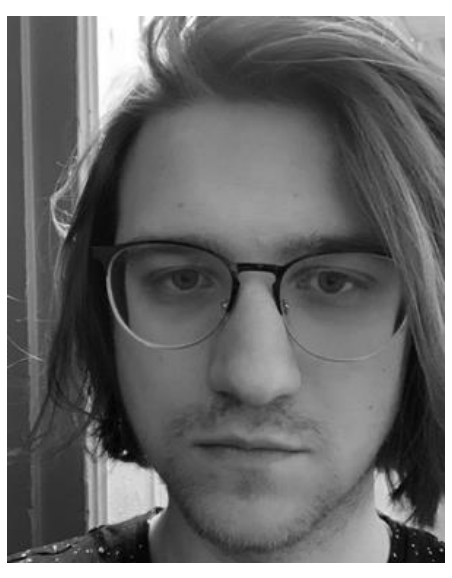

**Georgiy A. Solomakha** was born in Murmansk, Russia, in 1994. He received the M.Sc. degree in radiophysics from St. Petersburg Polytechnic State University, Saint Petersburg, Russia, in 2017, and the Ph.D. degree in antennas and microwave devices from the School of Physics and Engineering, ITMO University, Saint Petersburg, in 2021. He is currently a Post-Doctoral Researcher with the School of Physics and Engineering, ITMO University. His current research interests include magnetic resonance imaging (MRI) coils for ultrahigh-field MRI.